# Astrophysics:   A burst of new ideas

Bing Zhang

**Gigantic cosmological γ-ray bursts have fallen into a dichotomy of long and short bursts, each with a very different origin. The discovery of an oddball burst calls for a rethink of that classification.**

The events known as γ-ray bursts (GRBs) are the most violent and luminous explosions observed in the Universe. In the early 1990s, it became clear that they come in two distinct flavours: longer-duration bursts, typically longer than 2 seconds, with a spectrum of emitted radiation that peaks at lower ('softer') energy; and shorter-duration bursts with a more energetic, 'harder' spectrum[1]. Observations of burst afterglows in the past decade — particularly in the last year[2,3,4] — have seemed to show that this division is a clean one, and is firmly rooted in the progenitor of each type of burst. According to this picture, long bursts are associated with a young stellar population, marking the deaths of massive stars whose lifetime is short[5]. Short bursts, on the other hand, are associated with an old stellar population, and are probably powered by mergers of compact objects such as neutron stars or black holes[6].

In this issue, four papers[7–10] blow a hole in this cosy paradigm. They contain observations of a bright γ-ray burst, GRB 060614, that triggered NASA's GRB sentinel, the Swift satellite, at 12:43:48 UT on 14 June 2006. The burst defies pigeonholing within the current scheme.

Gehrels *et al.* (page 1044)[7] detail the circumstances of this peculiar burst's discovery. It is one of the brightest bursts ever seen, and was soon located precisely not only by Swift's instrumentation, but also by other space- and ground-based telescopes. The burst is situated in the suburbs of a faint and relatively nearby dwarf galaxy[8]. Its duration, recorded by Swift as 102 seconds[7], characterizes it unambiguously as a long GRB. According to previous experience, evidence for the death of a star in a stellar explosion – a supernova – should have been spotted in the burst's neighbourhood before too long. But the many optical telescopes around the world trained on the target, waiting for yet another confirmation of the connection between GRBs and supernovae, saw nothing.

Three further papers[8–10] provide independent report of the stringent upper limits on the radiation flux from a possible supernova underlying GRB 060614. Gal-Yam and colleagues' (page 1053)[8] made a series of observations with the Hubble Space Telescope in the weeks after the burst trigger. These set an upper limit more than 100 times fainter than the faintest supernova previously associated with a GRB — and indeed considerably fainter than any supernova ever observed. Della Valle *et al.* (page 1050)[9] report complementary observations from the European Southern Observatory's Very Large Telescope in the Atacama desert in northern Chile. This survey started 15 hours and ended 65 days after the burst, and provides an upper limit on the flux that is about three times higher than that of Gal-Yam and colleagues', but still well below the luminosity of any known supernova over an unprecedented long span of time. Fynbo *et al.* (page 1047)[10] use a range of telescopes to arrive at a similar result — and also discover a second long burst with no apparent supernova signature.

The absence of a supernova need not in itself be revolutionary. The production of a significant amount of nickel-56, which is a prerequisite for a supernova, is not guaranteed in a collapsing star[7,8], and the earliest model to connect GRBs with the massive-star collapses indeed characterized the bursts as 'failed supernovae'[5]. A supernova might also precede its associated GRB[11]. Nevertheless, the weight of evidence from the past decade is consistent with there being no significant gap between a GRB and its supernova, as well as with the hypothesis that every long GRB has a supernova accompanying it[12].

What makes the story more intriguing is that every property of GRB 060614 places it in the short burst category — except, that is, for its duration. Gehrel *et al.*[7] show that the time-lag of softer radiation behind harder parts of the burst's spectrum is small, as it is in a short GRB. Gal-Yam *et al.*[8] find that the afterglow of the burst has a large offset from the star-forming region of the host galaxy; again, a feature more characteristic of a short GRB. Similarly, the star-forming rate of the host galaxy is relatively small compared with those of normal long GRBs[8–10], consistent with an old stellar population more likely to host a short burst.

Even regarding its duration, the seemingly long GRB 060614 can be shortened. Its γ-ray light curve consists of a short, hard early episode lasting around 5 seconds followed by a long, soft tail[7]. Recent observations also indicate that most 'short' GRBs are not necessarily so short, and are usually followed by a softer emission tail lasting around 100 seconds[3,4]. Using an empirical relation between the spectrum hardness and the total energy budget of GRBs, it is possible to show[13] that GRB 060614 would be marginally classified as short if only it were around eight times less energetic.

So how can this burst and the classification scheme be squared? There are in principle three possibilities[8]. First, GRB 060614 is indeed a long GRB associated with a collapsing star. If so, its progenitor must be very different from those of most other long GRBs because of the anomalous properties detailed above. Second, the burst belongs to the merger-type short GRBs — in which case, these should not carry the name 'short' any more. Third, this is the prototype of a completely different, third category of bursts.

Given that the long–short paradigm is no longer adequate to describe the entire GRB phenomenon, a new terminology can be invented. Dividing bursts into Type I and Type II bursts by analogy with the supernova classification scheme might seem to lack imagination, but a comparison shows that such a definition might not be a bad choice (Figs 1,2)[13]. The progenitors of Type Ia supernovae, like those of the traditional short GRBs, belong to the old stellar population and live in binary systems. Similarly, the progenitors of type II supernovae, like those of traditional long GRBs, are collapsing massive stars who die at a young age.

Comparing the observational properties of GRB 060614 (shaded cells)[7–10] with the multiple criteria in Figure 2, one can see that this burst merges the properties of both categories. If one prefers to fit this burst into the straightjacket of bimodal classification, as I do, it would seem safer to apportion it to Type I, the traditional short category. There are great theoretical difficulties in producing extended radiation emission from a merger of compact stars, although various ideas to overcome these problems have been suggested[4,14–17].

To resolve definitively whether GRB 060614 is a peculiar example of a Type I burst, or whether it is a representative of a third class of object, more data are needed. In particular, given that our current information about the make-up of this GRB's host galaxy does not fully rule out a Type II origin, discovering whether or not GRBs with similar properties will be detectable in elliptical host galaxies[8,13], which consist entirely of older stellar populations, will hold the key to the final answer.

**Bing Zhang is in the Department of Physics and Astronomy, University of Nevada, 4505 Maryland Parkway, Las Vegas, Nevada 89154-4002, USA.**
**e-mail: bzhang@physics.unlv.edu**

**Figure 1 The classification of supernovae[18].**

|  | Type Ia | Type II (Type Ib/c) |
|---|---|---|
| Stellar population | Old | Young |
| Host galaxy | All types of galaxies | Late-type galaxies |
| Progenitor | Binary systems (accreting-induced collapse of white dwarfs) | Single star systems (core collapses of massive stars) |

**Figure 2 A classification scheme for γ-ray bursts.** The yellow rows show the analogy with the supernova classification scheme. Shaded cells show the properties of GRB 060614 (refs 7–10).

|  | Type I (short–hard) | Type II (long–soft) |
|---|---|---|
| Duration | Usually short (may have a long tail?) | Usually long |
| Spectrum | Usually hard (tail is soft) | Usually soft |
| Spectral lag | Short | Long |
| SN association | No | Yes |
| Stellar population | Old population | Young population |
| Host galaxy | All types of galaxies (predominantly in regions of low star formation rate) | Late-type galaxies (predominantly in irregular, dwarf galaxies) |
| Location in the host galaxy | Outskirts | Central |
| Progenitor | Mergers of compact objects in binary systems? | Single star systems? (core collapse of massive stars) |